\documentclass{aa}
\usepackage{graphicx}
\begin{document}
\title{Extended Star Formation in  
Dwarf Spheroidal Galaxies: the Cases of Draco, Sextans, 
and Ursa Minor}

\titlerunning{Extended Star Formation in Satellite Dwarf Galaxies}
\authorrunning{C. Ikuta \& N. Arimoto}

   \author{C. Ikuta 
          \inst{1}\fnmsep\thanks{JSPS fellow}
          \and
          N. Arimoto\inst{2,3}
          }

   \offprints{C. Ikuta}

   \institute{School of Physics \& Astronomy,      
         University of Nottingham, Nottingham, NG7 2RD, U.K.\\
              \email{chisato.ikuta@nottingham.ac.uk}
         \and
        National Astronomical Observatory of Japan,
              2-21-1, Osawa, Mitaka, Tokyo 181-8588,  
              JAPAN\\   
             \email{arimoto@optik.mtk.nao.ac.jp}
        \and
        Institute of Astronomy, University of Tokyo,
            2-21-1, Osawa, Mitaka, Tokyo 181-0015, 
           JAPAN\\
             }

   \date{Received 18 October 2001; accepted 19 April 2002}

\abstract{
Star formation and chemical enrichment histories 
of the dwarf spheroidal galaxies (dSphs) Draco, Sextans, 
and Ursa Minor are investigated by means of chemical evolution models and 
a simulation code for colour-magnitude diagrams (CMDs). 
The CMD simulation code is designed to 
fully consider effects of the chemical evolution on  
stellar evolution and photometric properties. 
For this aim, star formation and chemical enrichment histories 
are calculated consistently in the code. 
Comparisons between the chemical evolution models and 
the observed abundance patterns reveal that 
the star formation rates were very low ($1-5$\,\% of that 
of the solar neighbourhood disc) and that 
the initial star formation continued 
for a long duration ($> 3.9-6.5$ Gyr) in these dSphs. 
This star formation history can reproduce 
morphologies of the observed CMDs, 
such as narrow red giant branches and 
red horizontal branches and succeeds in
solving the second parameter problem of the dSph Draco. 
Hence, both of the abundance patterns 
and the morphologies of the CMDs 
can be explained by the star formation histories characterised 
by the low star formation rate and the long duration of the 
star formation period. Because of the low star formation rates, 
plenty of gas remains at the final epoch 
of star formation. We suggest that gas stripping by the Galaxy results in 
termination of star formation in the dSphs. 
}

\maketitle

\section{Introduction}
Satellite dwarf spheroidal galaxies of the Milky Way 
have been considered to be similar to globular clusters (e.g., Hodge 1971), since 
they are composed of old metal-poor stars and contain little gas.  
Recent analyses of the colour-magnitude diagrams (CMDs), 
however, reveal a surprising fact that dwarf spheroidal 
galaxies (dSphs) are of composite stellar populations. 
Some of dSphs formed stars continuously and/or intermittently 
for many Gyr (e.g., the dSph Carina: Smecker-Hane et al. 1996; 
the dSph Leo I: Gallart et al. 1999). 
Low-resolution spectra of individual 
stars along the red giant branch (RGB) have shown that both 
Galactic and M31 satellite dSphs have internal metallicity dispersions 
(e.g., Suntzeff et al. 1993; Smecker-Hane \& McWilliam 1999; 
C$\hat{\rm o}$t$\acute{\rm e}$, Oke, \& Cohen 1999). 
Using high dispersion (R=34000) spectroscopy, 
Shetrone, C$\hat{\rm o}$t$\acute{\rm e}$, \& Sargent (2001) have recently
presented abundances of various elements in 
the three dSphs Draco, Sextans, and Ursa Minor,
which are all satellite galaxies of the Milky Way.
They report that the abundance ratios of $\alpha$ 
elements (O, Si, Mg, etc.) 
relative to iron ([$\alpha$/Fe]) are nearly equal to solar 
at low metallicity ([Fe/H] $< -1.4$). 
This suggests that the stars were formed in the gas enriched  
by Type Ia supernovae (SNe\,Ia)  as well as by 
Type II supernovae (SNe\,II),  
that is, star formation in the dSphs continued over 
the lifetime (typically $1-2$ Gyr) of SNe\,Ia progenitors. 
This result allows us to directly study the 
chemical enrichment histories (CEHs) of 
the dSphs Draco, Sextans, and Ursa Minor.

One problem in deriving an accurate SFH 
from stellar colours is the so-called age-metallicity degeneracy. 
Older stars become redder and 
the same is true for more metal-rich stars. 
Hence, young and metal-rich stars show colours similar 
to old and metal-poor ones. 
We have empirically solved the 
age-metallicity degeneracy
for the three dSphs cited above to derive  
their star formation histories (SFHs) 
to a level of accuracy which is impossible 
in other galaxies in the Local Group.  
To reveal SFHs in detail would help to understand how large galaxies 
and the environments affect the evolution of dwarf galaxies.

In addition to a study of the evolution of dwarf galaxies themselves, 
one can also learn about the possible merging history of 
the Milky Way by comparing the abundances of the satellite 
dwarf galaxies with those of Galactic field halo stars. 
In a hierarchical galaxy formation scenario, 
larger galaxies grew at the expense of  
smaller gaseous fragments (e.g., White \& Rees 1978; Kauffmann, 
White \& Guiderdoni 1993; Blumenthal et al. 1994; 
Cole et al. 1994). From the observational perspective, 
Searle \& Zinn (1978) proposed 
a model in which the Galactic halo was formed {\it via} 
infall and destruction of proto-Galactic fragments.  
Increasing empirical and analytical evidence 
suggests that the Galactic halo was, at least in part, 
assembled from chemically-distinct, low-mass fragments 
(e.g., Searle \& Zinn 1978; Yanny et al. 2000). 
Such evidence includes the recent 
discovery of the tidally disturbed dSph 
Sagittarius which is falling onto 
the Galaxy (Ibata, Gilmore \& Irwin 1994)
as well as the numerous reports of kinematical substructure among 
halo field stars (e.g., Chen 1998). 
The scenario can be tested by comparing abundance 
patterns of field stars in the Galactic halo 
and those of satellite dwarf galaxies. 
If today's dwarf galaxies in the Local Group 
were the counterparts of proto-Galactic gaseous fragments, 
the abundance patterns of dwarf galaxies should be similar to those 
of field stars in the Galactic halo.
The observations by Shetrone et al. (2001), however, have shown that the 
patterns of abundance ratio in the dSphs are different 
from those of the Galactic halo stars. 
The dSphs have $-0.4 \le {\rm [Mg/Fe]} \le 0.4$, 
while field stars in the Galactic halo have 
$0.2 \le {\rm [Mg/Fe]} \le 0.6$ . 
The authors concluded that the Galactic halo was unlikely 
to have formed {\it via} 
the accretion of objects similar to the dSphs Draco, Sextans, and Ursa Minor.

In this paper, we discuss the SFHs and the CEHs of 
the three dSphs Draco, Sextans, and Ursa Minor. 
Using the abundance patterns and the CMDs, 
we show that the low (1$-5$\,\% of the solar neighbourhood) 
star formation rate and relatively long duration ($3.9-6.5$ Gyr) 
of initial star formation explain both the abundance patterns and 
morphologies of CMDs such as well-populated 
red horizontal branch (RHB) and the tight red giant branch (RGB).

The paper is organised as follows. 
In section\,2, chemical evolution model is confronted with the observed stellar abundances 
to study the CEHs and SFHs. 
In section\,3, by calculating CMDs by using our CMD simulation code 
(Ikuta 2001), we will show that the SFHs derived 
in the previous section succeed in reproducing observed CMDs. 
Discussions and conclusions are presented 
in sections\, 4 and 5, respectively.

\section{Observed Abundances and Chemical Evolution}
\subsection{Chemical evolution model}

\begin{figure}
\centering
\resizebox{9cm}{!}{\includegraphics{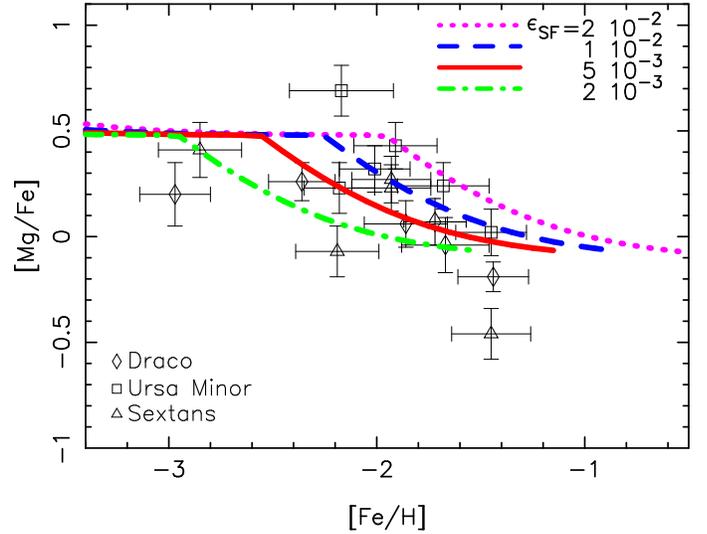}}
\caption
{Theoretical abundance patterns together with the observations 
(Shetrone et al. 2001)  for stars in the dSphs Draco, Sextans, 
and Ursa Minor.  The meaning of the marks is written on the panel.
The same IMF (the Salpeter IMF: $x=1.35$) 
with upper ($m_u=60M_\odot$) and lower ($m_l=0.1M_\odot$) 
mass limits are adopted in all the model, while 
different values of $\epsilon_{\rm SF}$\,Gyr$^{-1}$) 
are assumed; model\,A ($\epsilon_{\rm SF}=2\,10^{-2}$\,Gyr$^{-1}$: 
dotted line); model\,B ($\epsilon_{\rm SF}=1\,10^{-2}$\,Gyr$^{-1}$:   
dashed line); model\,C ($\epsilon_{\rm SF}=5\,10^{-3}$\,Gyr$^{-1}$: 
solid line); and model\,D ($\epsilon_{\rm SF}=2\,10^{-3}$\,Gyr$^{-1}$: 
dotted-dashed line). Colour version of this figure 
appears in the electric journal. 
}
\end{figure}

Generally stellar birth rate is separated into two
independent functions. The birth rate of stars with mass between 
$m$ and $m+dm$ is described as $C(t)\phi(m)dm$,
where $C(t)$ and $\phi(m)$ are the SFR and the 
initial mass function (IMF), respectively (Tinsley 1980).
The IMF is assumed to be time invariant 
with a power-law spectrum. Normalising to unity, we have
\begin{equation}
\phi(m)=\frac{(x-1)m_l^{x-1}}{1-(m_l/m_u)^{x-1}} m^{-x},
        \label{eqn:imf1}
\end{equation}
where the lower and the upper mass limits are assumed to be 
$m_l=0.1M_\odot$ and $m_u=60M_\odot$,
respectively. The Salpeter IMF has $x=1.35$ in this definition. 
Similarly to the solar neighbourhood disc model of 
chemical evolution (Arimoto, Yoshii, \& Takahara 1992), 
the Salpeter IMF is assumed in this section.
Models with different IMFs will be discussed in section 4. 
The gas mass $M_{\rm g}(t)$ changes through star formation:
\begin{equation}
\frac{dM_{\rm g}(t)}{dt}=-C(t)+E(t),
\end{equation}
where $E(t)$ is the gas ejection rate from dying stars, 
assuming a closed-box model. 
In the case of the Galactic disc, an 
infall model is generally assumed 
to solve the so-called G-dwarf problem (e.g., Tinsley 1980; 
Arimoto et al. 1992). 
However, there is no observational evidence for the G-dwarf problem 
in dwarf galaxies, thus we do not consider the infall of gas here. 
Gas outflow during the star formation is not considered, either. 
A recent numerical experiment showed that 
the fraction of mass loss from dwarf galaxies is small 
(at most $\sim 7$ percent of the galaxy mass; MacLow \& Ferrara 1999). 
Metal-enhanced wind is often introduced to 
explain observed properties of dwarf galaxies (e.g., Vader 1986). 
MacLow \& Ferrara (1999) showed that indeed most of the metals are able to 
leave dwarf galaxies during star burst, 
although the mass loss rate is small. 
Virtually no metals are retained in dwarf galaxies 
if high energy is supplied by supernova explosions 
(see table\,3 in their literature for details).
Complete metal loss means no chemical enrichment in dwarf galaxies. 
This is inconsistent with observations which show clearly that dSphs are
chemically polluted. Their results suggest that 
only star formation of a low rate
allows a dwarf galaxy to retain a large fraction of heavy
elements produced in massive stars.  
As will be shown later, very low SFRs are indeed 
required to explain the observed abundances and CMDs. 
Thus, a simple model is adopted here and a galaxy is assumed to be
a closed system.

\begin{table}
\caption[]{Model Parameters}
\begin{tabular}{c|ccc}
\hline
\hline
Model  & $x$ & $\epsilon_{\rm SF}$ (Gyr$^{-1}$) & $\Delta T_{\rm SF}$ (Gyr) \\
\hline
A & 1.35 & $2 \cdot 10^{-2}$ & 2.5 \\
B & 1.35 & $1 \cdot 10^{-2}$ & 3.9 \\
C & 1.35 & $5 \cdot 10^{-3}$ & 6.5 \\
D & 1.35 & $2 \cdot 10^{-3}$ & 12  \\
E & 1.75 & $2\cdot 10^{-1}$  & 1.6 \\
F & 1.95 & $2\cdot 10^{-1}$  & 1.8 \\
G & 2.15 & $2\cdot 10^{-1}$  & 2.2 \\
\hline
\end{tabular}
\end{table}

\begin{figure*}
\vspace*{1cm}
\begin{center}
\setlength{\unitlength}{0.8mm}
\begin{picture}(120,220)

\put(10,220){\thicklines\framebox(120,10){%
(1) Assumption of SFH ($\epsilon_{\rm SF}, \Delta T_{\rm SF}$, %
Age, IMF)}}
\put(70,220){\vector(0,-1){8}}

\put(10,200){\thicklines\framebox(120,10){%
(2) Calculation of CEH}}
\put(70,200){\vector(0,-1){8}}

\put(10,180){\thicklines\framebox(120,10){%
(3) Calculation of Isochrones}}
\put(70,180){\vector(0,-1){8}}

\put(10,160){\thicklines\framebox(120,10){%
 Monte Carlo Simulation}}
\put(70,160){\vector(0,-1){8}}

\put(10,140){\thicklines\framebox(120,10){%
(4) HR Diagram Simulation}}
\put(70,140){\vector(0,-1){8}}

\put(10,120){\thicklines\framebox(120,10){%
(5) Model Atmosphere}}
\put(70,120){\vector(0,-1){8}}

\put(10,100){\thicklines\framebox(120,10){%
(6) Colour-Magnitude Diagram}}
\put(70,100){\vector(0,-1){8}}

\put(10,80){\thicklines\framebox(120,10){%
(7) Observational Errors and Data Incompleteness}}
\put(70,80){\vector(0,-1){8}}

\put(10,60){\thicklines\framebox(120,10){%
Monte Carlo Simulation}}
\put(70,60){\vector(0,-1){8}}

\put(10,40){\thicklines\framebox(120,10){%
(8) Colour-magnitude Diagram}}
\put(70,40){\vector(0,-1){8}}

\put(130,25){\vector(1,0){8}}
\put(140,20){\thicklines\framebox(20,10){%
No}}

\put(10,20){\thicklines\framebox(120,10){%
(9) Comparison with Observation}}
\put(70,20){\vector(0,-1){8}}

\put(10,0){\thicklines\framebox(120,10){%
(10) SFH \& CEH}}

\put(150,30){\vector(0,1){195}}
\put(150,225){\vector(-1,0){15}}
\end{picture}
\caption{Flow-chart of a CMD simulation code.}
\end{center}
\end{figure*}
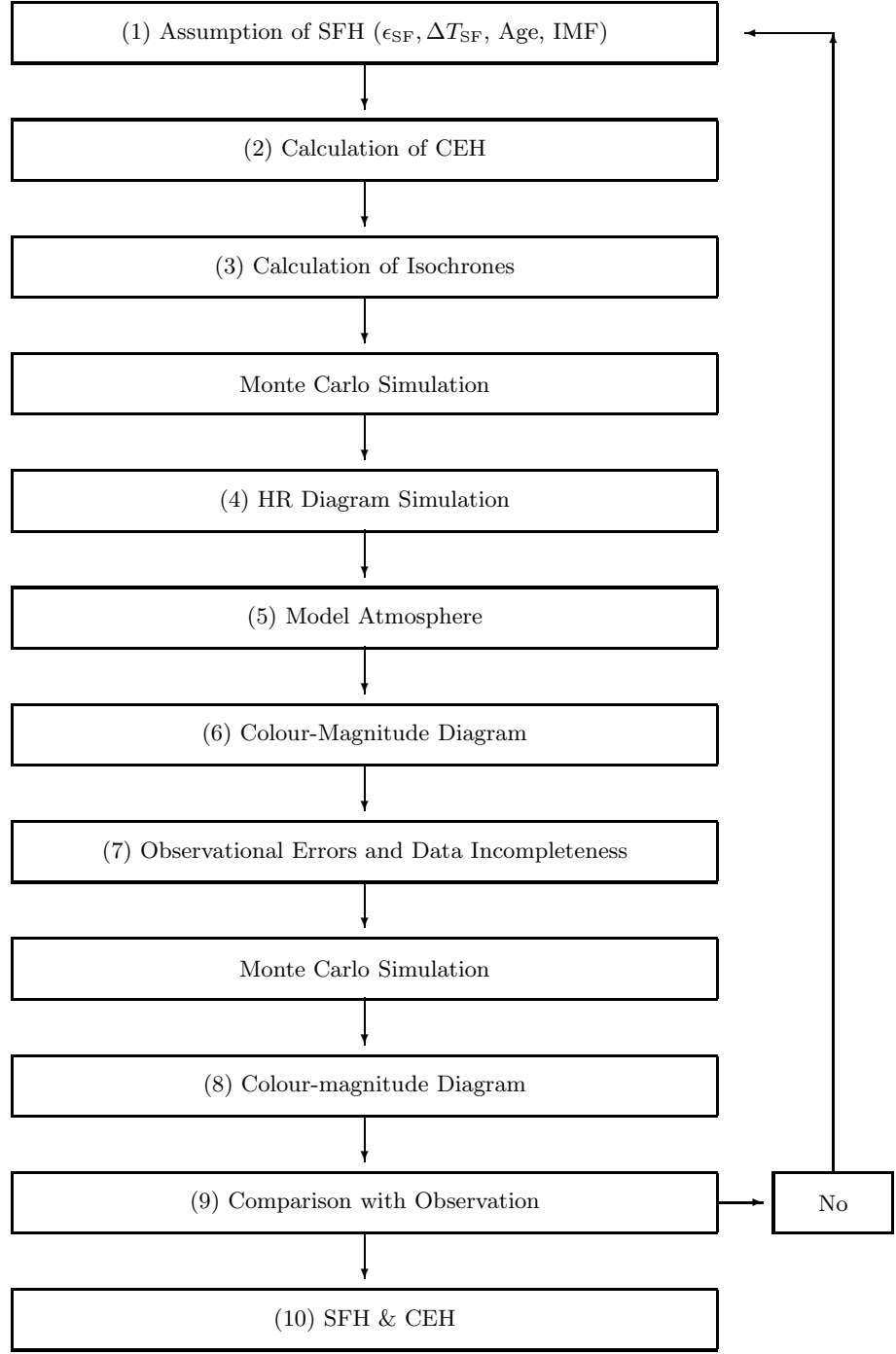

The SFR is assumed to be proportional 
to the gas fraction $f_{\rm g}(t)$:
\begin{eqnarray}
C(t) =\epsilon_{\rm SF} M_{\rm G}\cdot f_{\rm g}(t) 
= \epsilon_{\rm SF}\cdot M_{\rm g}(t)
\end{eqnarray}
with
\begin{equation}
f_{\rm g}(t)\equiv M_{{\rm g}}(t)/M_{\rm G},
\end{equation} 
where $\epsilon_{\rm SF}^{-1}$ (Gyr) 
and $M_{\rm G}$ are the timescale of star formation
and the initial total mass of a galaxy, respectively. 

The evolution of the abundance 
of the $i$-th element $Z_i(t)$ is given as 
\begin{equation}
\frac{d(Z_iM_{\rm g})}{dt}=-Z_i C(t)+E_{Z_i}(t)
+Z_i(0)M_{\rm G},
\end{equation}
where $E_{Z_i}(t)$ is the total ejection rate of
processed and unprocessed $i$-th element,  
$Z_i(0)$ is the initial abundance of $i$-th element in
a proto-cloud. Hereafter, $Z_i(0)=0$ is always assumed, 
i.e., a proto-cloud was metal-free.
We calculate $E(t)$ and $E_{Z_i}(t)$ from the following integrals:
\begin{equation}
E(t)= \int_{m_t}^{m_u}(1-w_m) C(t-t_m) \phi(m)dm ,
\end{equation}
\begin{eqnarray}
\lefteqn{E_{Z_i}(t)} \nonumber\\
&&=\int_{m_t}^{m_u}\!\!\!\!\!\!\left[{(1\!-\!w_m)Z_i(t-t_m)\!+\!p_{im}}\right] 
C(t-t_m) \phi(m)dm ,
\end{eqnarray}
where $t_m$ is the lifetime of a star with mass $m$ and  
the lower limit
$m_t$ is the stellar mass with lifetime $t_m=t$. 
Nucleosynthesis data $w_m$ and $p_{im}$ are the remnant mass fraction
and the total mass fraction of processed and unprocessed
$i$-th element. 
It is still debated whether stellar metallicity 
affects both $w_m$ and $p_{im}$,
but no firm consensus has been reached yet.
Thus we neglect these possible effects:
instead we adopt the values calculated for the solar abundance.
We assume that the lower mass limit $m_{l,{\rm II}}$
for a progenitor of Type II supernova (SNeII)
is $10M_\odot$ (Tsujimoto et al. 1995).
Nucleosynthesis of SNeII and 
of SNeIa are explicitly included, since the 
former profoundly produce the  $\alpha$ elements 
(e.g., O, Mg, Si, Ca), 
while the latter the iron-peak elements (e.g., Fe, Ni, Co).
The data of nucleosynthesis are taken 
from Thielemann, Nomoto, \& Hashimoto (1996) 
for SNeII, Nomoto, Thielemann, \& Yokoi (1996) for SNeIa, 
and Renzini \& Voli (1981) for low-mass ($\le 8\,M_\odot$) stars. 
Progenitors of SNe\,Ia are generally considered to have 
longer lifetimes than those of SNe\,II, since a majority of them are 
likely to be Chandrasekhar mass white dwarfs 
(e.g., Nomoto, Iwamoto, \& Kishimoto 1997). 
Observations for disc (Edvardsson et al. 1993)  
and halo (e.g., McWilliams et al. 1995; Gratton \& Sneden 1987) 
stars in the Milky Way  show that there is a sudden decline of [$\alpha$/Fe] 
in the [$\alpha$/Fe]-[Fe/H] diagram. The decline  
is generally interpreted as an onset of SN\,Ia explosions. 
Studying observed [O/Fe] break 
at [Fe/H] $\sim -1$ for disc and halo stars in the Milky Way, 
Yoshii, Tsujimoto, \& Nomoto (1996) 
suggested that the effective lifetime of SNe\,Ia 
is around $1.5$\,Gyr.  A new SNIa model (Hachisu, Kato, \& Nomoto 1999) 
suggested that occurrence of SNeIa depends on the metallicities of progenitors 
as well as the lifetimes. Based on the new SN\,Ia model, 
Kobayashi et al. (1998) predicted that SN\,Ia events occur only 
for the progenitors with [Fe/H] $\ge -1.1$. On the observational side, 
$\alpha$ element to iron abundance ratios 
of the three dSphs, Draco, Sextans, and Ursa Minor 
(Shetrone et al. 2001) have indicated   
a significant contribution from SNeIa at a metallicity lower than 
[Fe/H]$=-1.0$.  The observed trends of [$\alpha$/Fe] 
are difficult to explain by the new SNIa model. 
Thus, the metallicity effect on SNeIa is not considered here 
and the lifetime of the progenitors 
is assumed to be $1.5$ Gyr (Yoshii et al. 1996). 

\begin{figure}
\resizebox{8cm}{!}{\includegraphics{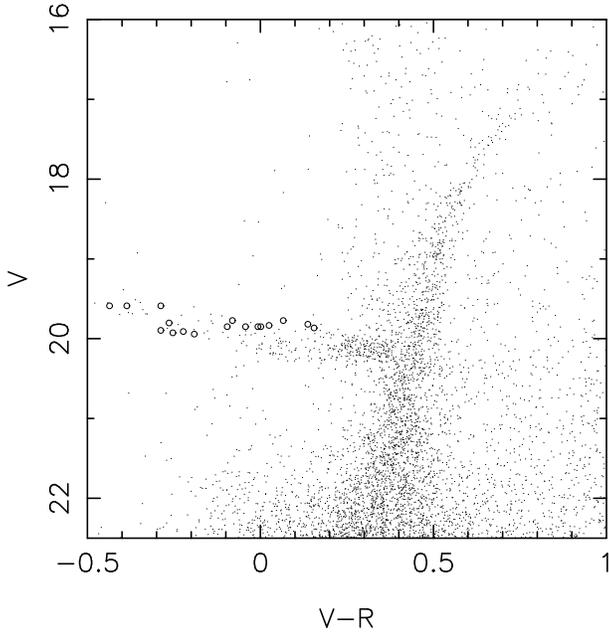}}
\caption{
CMD of the C0 region of the Draco dSph observed by 
Piatek et al. (2001). The points marked with open circles 
correspond to some confirmed RR Lyrae stars (Baade \& Swope 1961).
}
\end{figure}

\subsection{Comparison with the observation}

Fig.\,1 reproduces the observed stellar abundances of the dSphs 
Draco, Sextans, and Ursa Minor in the [Mg/Fe]$-$[Fe/H] diagram 
(Shetrone et al. 2001). Despite the narrow widths of the RGB 
(see figure\,1 of Shetrone et al. 2001), the iron abundances exhibit 
a large dispersion ($-3 \le {\rm [Fe/H]} \le -1.4$), and
the abundance ratios of the $\alpha$-elements 
(e.g., Mg) to iron are near or below solar. 
The abundance ratio [Mg/Fe] starts to decrease 
at [Fe/H] $\simeq -2$. 
On the other hand, [Mg/Fe] in the solar neighbourhood 
decreases with [Fe/H] at around [Fe/H]$\ge -1$, 
which is generally interpreted as 
the onset of SNeIa explosions (e.g., Matteucci \& Greggio 1986).  
Supposing that the decline of 
[Mg/Fe] in the dSphs is also caused by SNeIa, the star formation 
should have continued longer than the lifetime 
(typically $1-2$\,Gyr) of progenitors of SNeIa and the SFRs 
should be much lower than that of the solar neighbourhood.  
A time-delay model (Matteucci \& Brocato 1990) 
of the SNeIa precisely predicted that the low star formation rate 
results in lower [Mg/Fe] ratios relative to stars in the Galaxy.

In Fig.\,1, theoretical abundance patterns of our models A$-$D 
are superposed. The models assume different $\epsilon_{\rm SF}$ 
of $2\cdot10^{-2}, 1 \cdot 10^{-2}, 
2\cdot 10^{-3}, 1 \cdot 10^{-3}$ (Gyr$^{-1}$) as indicated in Fig.\,1. 
The model parameters are summarised in table\,1. 
A typical value of $\epsilon_{\rm SF}$ for the solar neighbourhood disc is 
$\epsilon_{\rm SF}=0.2$\,Gyr$^{-1}$ (Arimoto et al. 1992). 

The models B$-$D shown in Fig.\,1  
give a good fit to the observed abundance patterns. 
Since the abundance ratios resulting from the model A are too 
 high to reproduce the observed ones, the model A is rejected.  
The model D with $\epsilon_{\rm SF}=1\cdot10^{-3}$ Gyr$^{-1}$  
is not appropriate, either.  Since 
chemical evolution is so slow, 
it takes around 15 Gyr to reach [Fe/H]$=-1.4$, 
which is the highest abundance observed in these galaxies. 
If this model is adopted, the star formation must continue 
till now. Clearly this is inconsistent with 
the fact that no star formation occurs currently in these galaxies. 
Thus, only models B and C remain as being acceptable. 
We note that the observational data for the dSph Sextans have 
lower signal-to-noise ratios than those for the dSphs Draco and 
Ursa Minor (see table\,2 of Shetrone et al. 2001) and  
the abundances of the dSph Sextans are less reliable than  
those of the other two dSphs. Thus, it is 
not necessary to take the discrepancy between 
the model predictions and the abundances of the dSph Sextans too seriously.  

In short, the abundance patterns of the dSphs Draco, Sextans, 
and Ursa Minor all suggest
that the chemical enrichment occurred with SFRs which are much lower
than normal spiral galaxies (only $1-5\%$).

\section{CMDs}
\begin{figure*}
\rotatebox{270}{\resizebox{!}{18cm}{\includegraphics{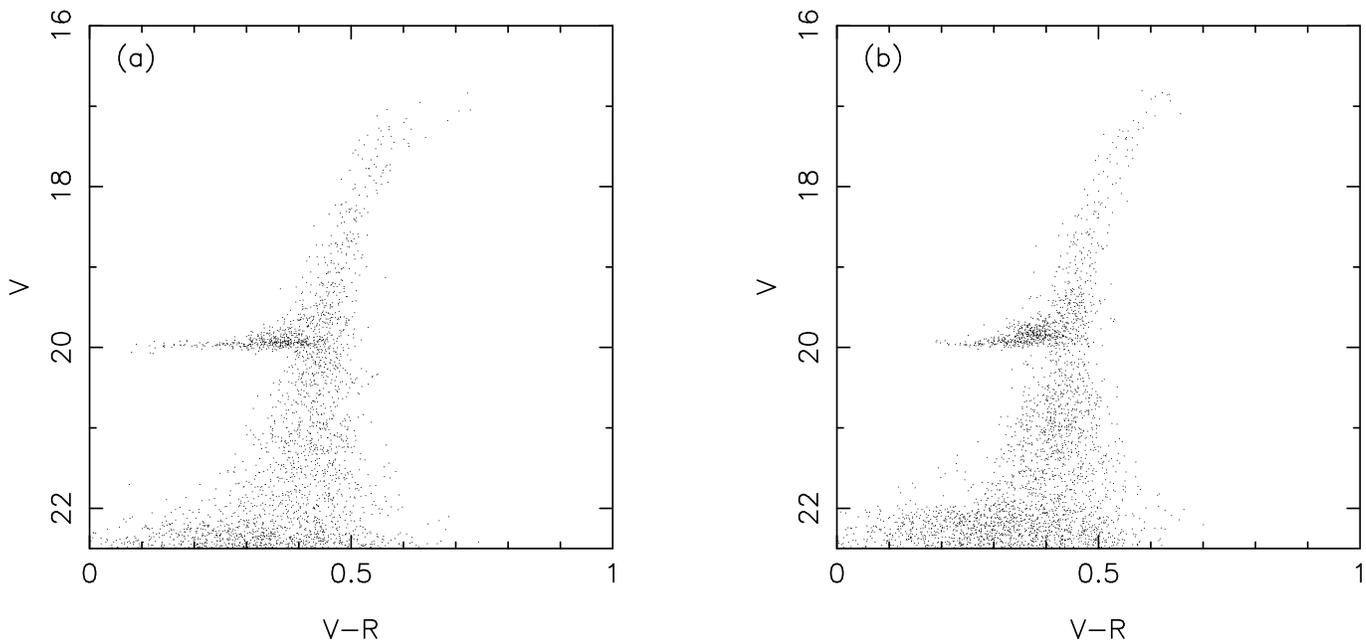}}}
\caption{
Simulated CMDs of models B (left) and C (right). 
In model\,B, $\epsilon_{\rm SF}=1\,10^{-2}$\,(Gyr$^{-1}$ and 
$\Delta T=3.9$\,Gyr are assumed, while 
$\epsilon_{\rm SF}=5\,10^{-3}$\,(Gyr$^{-1}$ 
and $\Delta T=6.5$\,Gyr are adopted in model\,C. 
In the both models, we assumed the same upper and lower mass limits 
($m_u=60M_\odot$ and $m_l=0.1M_\odot$, respectively), 
the Salpeter IMF, and the galactic age of 12\,Gyr. 
Observational errors and data incompleteness 
are taken from Piatek et al. (2001).
}
\end{figure*}

The models B and C fit well to the observed trends in the 
[Mg/Fe]$-$[Fe/H] diagram. Next, 
we discuss if the colour-magnitude diagrams (CMDs) predicted by 
these models can reproduce the observed ones. 
The CMDs provide us with the most 
detailed information to follow SFHs and CEHs back to the oldest stars. 
The SFH derived from the abundance patterns in the previous 
section must give a consistent CMD with the observations.

\subsection{Simulation code of a CMD}

Simulating numerical CMDs has become 
a standard technique to study the SFHs 
of nearby galaxies through the observed CMDs. 
Galaxies are composed of complex stellar populations,  
and the observed CMDs are affected by 
photometric errors and detection incompleteness. 
A Monte Carlo simulation allows us 
to take into account both simultaneously. 
A composite stellar population is randomly generated 
according to predictions of stellar evolution 
theory with the IMF and SFR being assumed, 
and then uncertainties of the data such as the increasing 
scatters and rising incompleteness 
at fainter magnitudes are taken into account.  

Because of these advantages, 
this approach is becoming more and more frequently  
used to study the stellar populations of nearby galaxies.  
The approach was first applied by Ferraro et al. (1989) 
and was fully described by Tosi (1991) and 
Greggio et al. (1993). A more quantitative method, 
the R-method, to compare simulated and observed CMDs 
was presented by Bertelli et al. (1992) and 
was described by Vallenari et al. (1996) in more detail.  
The statistical comparison was further developed 
by Tolstoy \& Saha (1996) and Tolstoy (1996) by adopting Bayes' theorem.    
Now several groups have constructed CMD simulators 
and have investigated SFHs of galaxies in the Local Group 
(e.g., Aparicio et al. 1996; Dolphin 1997; 
Hernandez, Valls-Gabaud, \& Gilmore 1999; Gallart et al. 1999).  

We should point out, however, that there were two problems in the previous studies. 
First, all the approaches so far developed 
use the so-called optimising 
method of stellar population synthesis. 
The best mixture of stellar population is searched iteratively 
to reproduce the CMDs. 
Problems with this approach are the unproven uniqueness of the solution 
(e.g., Greggio et al. 1999) and a lack of evolutionary information.

Second, stellar metallicity was assumed {\it a priori}
in all the simulation codes presented previously. 
Some introduced metallicity variation 
in time (e.g., Gallart et al. 1999), 
but the variation assumed was independent of the SFH. 
Since the metallicity affects evolutionary tracks 
and atmospheres of stars, 
the colours and luminosities of stars are changed by metallicity.   
Thus, it is crucial to use a simulator 
of CMDs which fully takes into account a CEH   
to interpret the CMDs properly and to derive the SFHs accurately. 
The simplification and/or neglect of chemical evolution lead to  
a serious problem called the age-metallicity degeneracy. 
Since stellar colours become bluer when stars are younger and/or poorer 
in metallicity, there are at least two interpretations 
for a given position 
of a star in the CMD; young and metal-rich or old and metal-poor. 
Therefore, an assumption of the metallicity (or age) 
of a certain stellar population may result in wrong estimation 
of age or metallicity, due to the age-metallicity degeneracy.

Aiming to solve the degeneracy, 
we have built a numerical simulation code of CMD morphology (Ikuta 2001). 
In the code, we adopt an evolutionary method 
of stellar population synthesis (e.g., Arimoto \& Yoshii 1986).   
A galaxy is assumed to have been a proto-galactic gas cloud 
at the beginning. 
Stars formed and newly processed elements were ejected from stars at 
the end of their lives either via stellar wind or 
via supernova explosions (SNeIa and SNeII).
The gas was chemically enriched 
and the next generation of stars changed their evolution 
and photometric properties due to a metallicity increase. 
The code is particularly designed to disentangle stellar age 
and metallicity both  of which heavily affect the morphology of CMDs.  
To take into account effects of chemical evolution, 
a fine interpolation in stellar ages and metallicities 
is adopted in our code.  
The later stages of stellar evolution, 
in particular, horizontal branch (HB),  
asymptotic giant branch (AGB),  and  post-AGB stars  
are treated in   a  sophisticated way, since these bright stars 
are  crucial for this purpose.
In this section, the procedure of our Monte Carlo simulation of CMDs 
and the ingredients are detailed. 

\begin{figure*}
\resizebox{17cm}{!}{\includegraphics{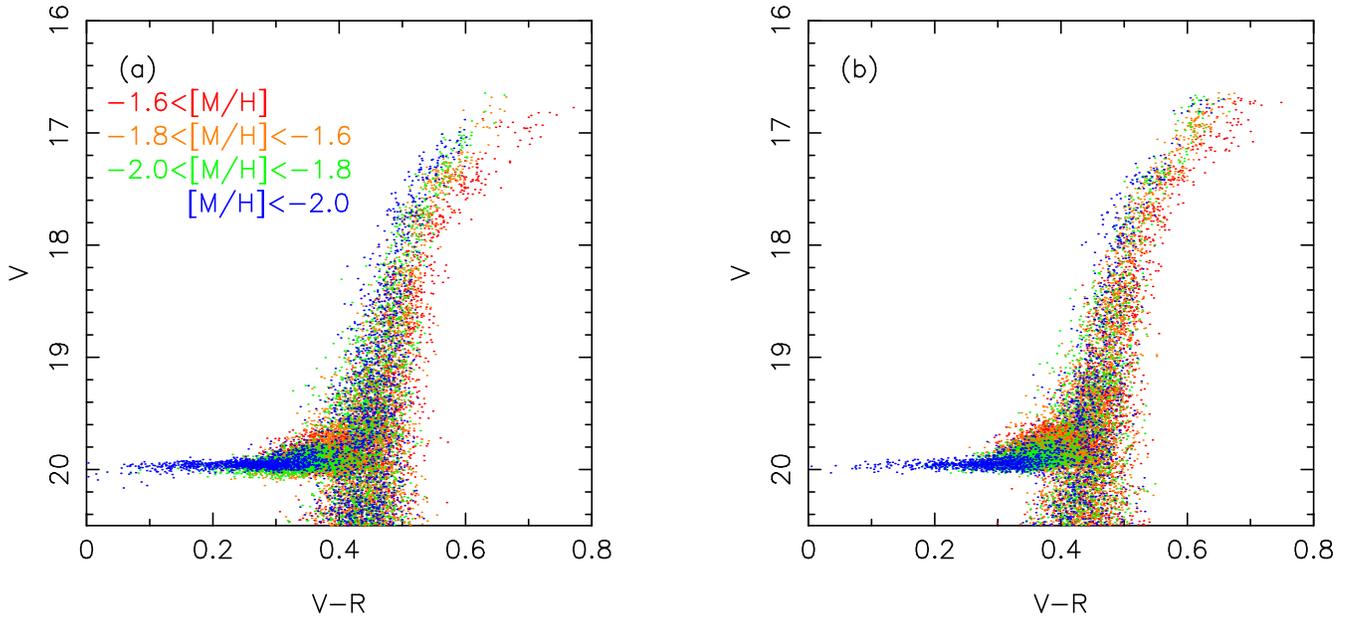}}
\caption{Simulated CMDs of models B (panel (a)) 
and C (panel (b)). Adopted 
errors and data completeness are the same as those in Fig.\,4, 
while the number of stars in each panel is increased for 
presentation purposes. 
Colours correspond to the following metallicity ranges;  
blue: [M/H]$<-2.0$; green: $-2.0 \le {\rm [M/H]} < -1.8$; 
orange: $=1.8 \le {\rm [M/H]} < -1.6$; red:$-1.6 \le$ [M/H]. }
\end{figure*}

\subsubsection{Outline of the Simulation Procedure}

Fig.\,2 shows the flow-chart of the simulation. 
First, a SFH is assumed. 
The parameters to describe the SFH are the 
timescale of star formation $\epsilon_{\rm SF}^{-1}$ (Gyr), 
the duration of star formation $\Delta T_{\rm SF}$ (Gyr), 
the slope of the IMF ($x$), and the age of a galaxy.  
Second, chemical evolution is calculated, which 
gives the total number of stars at each metallicity at each 
time step. Third, a HR diagram is calculated  
by using stellar evolutionary tracks, where a
Monte Carlo simulation is performed to assign stellar masses. 
Fourth, the effective temperature $T_{\rm eff}$ and 
luminosity $L$ of a star are transformed to 
colour and magnitude by using a library of stellar model atmospheres.
Fifth, a Monte Carlo simulation is performed again 
to take into account observational errors and detection completeness, 
and thus the CMD is obtained. 
A number of stars in the simulated CMD is adjusted to an observed CMD. 
Finally, the simulated CMD is confronted 
with the observed one by visual inspection.  If they are inconsistent, 
the parameters of the SFH are changed, and the simulation is re-started. 
If the simulated CMD fits the observation, 
the SFH and CEH are derived. Ingredients of the simulation, such as 
stellar evolutionary tracks, model atmospheres, 
and observational conditions, are described in the subsequent section.

\subsubsection{Stellar Tracks and Model Stellar Atmosphere}

        Stellar evolutionary tracks give luminosities, 
effective temperatures, and surface gravities of stars with 
given mass and chemical composition as a function 
of age. The tracks depend on the basic parameters 
such as the initial mass and chemical compositions. 
The properties of stellar populations in galaxies also 
depend on other properties not explicitly 
included in most of current stellar evolution models, i.e.,   
the stellar rotation and close binary companions. 
The numerical calculations are also sensitive to the treatment 
of the convection, such as a mixing length parameter 
$l/H_p$ and convective  overshooting.

The Padova stellar evolutionary tracks are adopted 
for stars from the main sequence (MS) 
to the early asymptotic giant branch (EAGB) stars. 
The Padova tracks cover wide ranges in age 
($0$ to $16$ Gyr) and metallicity ($Z=0.0001$ to $Z=0.05$), 
and are one of the most complete sets of stellar evolutionary 
models currently available. 
The Padova tracks were calculated with revised radiative 
opacities (Iglesias, Rogers, \& Wilson  1992) and with 
$l/H_p=1.63$. 
The adopted tracks were 
calculated for the following set of chemical compositions:  
($Y=0.230, Z=0.0001$; Girardi et al. 1996), 
($Y=0.230, Z=0.0004$; Fagotto et al. 1994a), 
($Y=0.240, Z=0.004$; Fagotto et al. 1994b),
($Y=0.250, Z=0.008$; Fagotto et al. 1994b), 
($Y=0.280, Z=0.02$; Bressan et al. 1993), and 
($Y=0.352, Z=0.05$; Fagotto 1994a).

To obtain the isochrone for a given metallicity, 
the original isochrones are linearly interpolated in metallicity.  
Since the metallicity changes stellar evolution non-linearly, 
any extrapolation in metallicity should  be avoided. 
Thus, stars of metallicity lower ($Z<0.0001$) and 
higher ($Z>0.05$) than the Padova tracks 
do not appear in our simulated CMDs 
and are not considered here. 
This is justified, because no significant populations of
such extreme metallicities are known to exist in dwarf galaxies.
In addition to these, close binary systems 
are not included, since the input tracks are only 
for isolated single stars.  
Although our code does not explicitly include binary stars, 
it effectively takes into account apparent 
binary stars as stellar blends.  Gallart et al. (1999) reported a study 
that is more detailed on this issue.

For the RGB evolution, the mass loss law of Reimers (1977) is adopted;
\begin{equation}
\dot{M}=-4 \cdot 10^{-13} \eta \frac{L/L_\odot}{g\cdot R/R_\odot}
~~~~~~(M_\odot \cdot {\rm yr}^{-1}),
\end{equation}
or equivalently,
\begin{eqnarray}
{\rm log}(-\dot{M}) =&-&4.67+{\rm log}\eta + 1.5{\rm log}(L/L_\odot) \nonumber\\
& & \hspace{1.8cm}-{\rm log}(M/M_\odot) -2{\rm log}T_{\rm eff},
\end{eqnarray}
where $L, g, R, T_{\rm eff},$ and $M$ are stellar luminosity, 
surface gravity, radius, effective temperature, and initial mass, respectively. 
The mass loss efficiency parameter 
$\eta$ is assumed to be $1/3$ (Renzini \& Voli 1981).

The horizontal branch (HB) stars are distributed on the HR diagram 
according to a modified Gaussian mass distribution 
equation (Lee et al. 1990);
\begin{eqnarray}
\Phi (M) &\propto& \left[ M-(\overline{M_{\rm HB}}-\Delta M)\right] 
(M_{\rm RGB} -M) \nonumber \\
& & \hspace{2.5cm} \times \exp \left[-\frac{(\overline{M_{\rm HB}}-M)^2}
{\sigma^2} \right], 
\end{eqnarray}
where $\sigma$ is a mass dispersion factor, 
and $\overline{M_{\rm HB}} \equiv (M_{\rm RGB}-\Delta M)$ 
is the mean mass of HB stars. A value of $\sigma=0.06M_\odot$ 
has been chosen to represent the HB distributions 
similar to those of Galactic globular clusters (Rood 1973; Lee et al. 1990). 
This is equivalent to considering the statistical dispersions 
of mass loss rate along the RGB.

To convert the theoretical temperature-luminosity data 
to the observable colour-magnitude plane, a stellar spectral library 
by Lejuene, Cuisinier \& Buser (1998) is used. 
This consists of Kurucz's (1992) spectra for hotter 
stars (O$-$K), Bessell et al.'s (1989; 1991) and Fruks 
et al's (1994) spectra for M giants, and Allard \& Hauschudt's (1995) 
for M dwarfs. In the original model spectra, 
systematic deviations become apparent 
when colour-temperature relations computed from the 
models are compared to the empirical ones at $1 Z_\odot$.  
The library adopted here is a version that 
the authors made by correcting the original library.  
The corrections are especially important for M star models. 
The fundamental parameters 
($T_{\rm eff}$, $g$, and [M/H]$=\log (Z/Z_\odot)$) 
are wide enough to cover all spectral types and luminosity 
classes that appear in observed CMDs; 
$3 \times 10^3 \le T_{\rm eff} \le 5 \times 10^4$,
$-1.02 \le \log g \le  5.5 $; 
$ -5.0 \le [{\rm M/H}] \le  1.0$. 
To reduce computational time, synthetic magnitudes and 
colours tabulated by Lejuene et al. (1998) are 
linearly interpolated in log$Z$, log$T_{\rm eff}$, and log$g$.

\subsection{Comparison with the CMD of the Draco dSph}

In section\,2, models B and C are shown to 
fit well the observed trends in the [Mg/Fe]$-$[Fe/H] diagram. 
Next, we discuss if the CMDs predicted by 
these models can reproduce the observed ones. 
Ideally, the SFH would be studied using a CMD  
which satisfies the following two conditions: (1) it covers a whole galaxy; 
(2) it is significantly deeper than the turnoff 
of the oldest stellar population.
Unfortunately, such CMDs are not available for the galaxies discussed here. 
The situation, however, is better for the dSph Draco.  
Recently Piatek et al. (2001) imaged nine fields in and around 
the Draco dSph by using the KPNO 0.9m telescope and presented 
the CMDs down to a luminosity level 
$\sim 2$ mag fainter than the HB.  
Thus, we mainly discuss the dSph Draco 
in this section and briefly mention the dSphs Sextans and Ursa Minor. 

Fig.\,3 shows the observed CMD 
of the C0 field in the Draco dSph (Piatek et al. 2001), which 
is characterised by the narrow RGB and the well-populated red HB. 
The HB in figure\,1 of Piatek et al. (2001) is somewhat distorted 
due to RR Lyrae stars which are marked with open circles in Fig.\,3. 
Therefore, we do not take seriously the inconsistency between 
the observed and simulated blue HB morphology (V$-$R$<0.1$). 
Historically, 
Baade \& Swope (1961) obtained the CMD of the Draco dSph and 
found that the RGB is generally similar to those 
of metal-poor globular clusters, although it is rather wide. 
The authors also found that the populous RHB  is incompatible 
with the low metallicity of a Population II system. 
This is the notorious second-parameter problem of the dSph Draco.  

\begin{figure}
\resizebox{8.5cm}{!}{\includegraphics{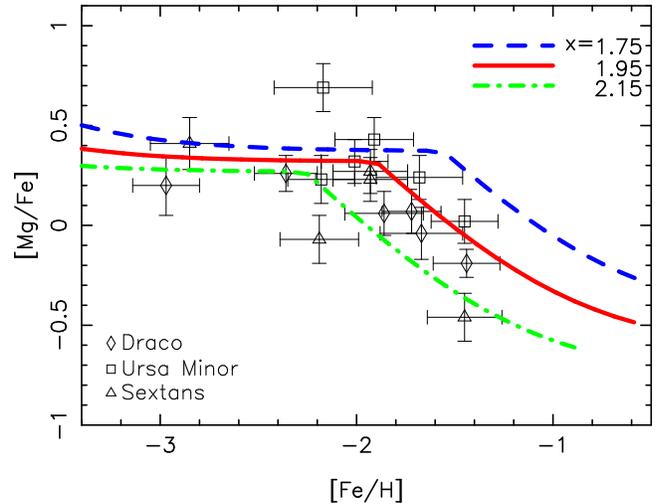}}
\caption{The same as Fig.\,1, but for models E-G.
Different lines mean different IMF powers; model\,E ($x=1.75$: dashed line);  
model\,F ($x=1.95$: solid line); model\,G ($x=2.15$; dotted-dashed line).  
The star formation time scale is the same as 
for the solar neighbourhood disc, i.e., 
$\epsilon_{\rm SF}=0.2$\,(Gyr$^{-1}$). See electric edition of the 
journal for a colour version of this figure.}
\end{figure}

\begin{figure*}
\rotatebox{270}{\resizebox{!}{18cm}{\includegraphics{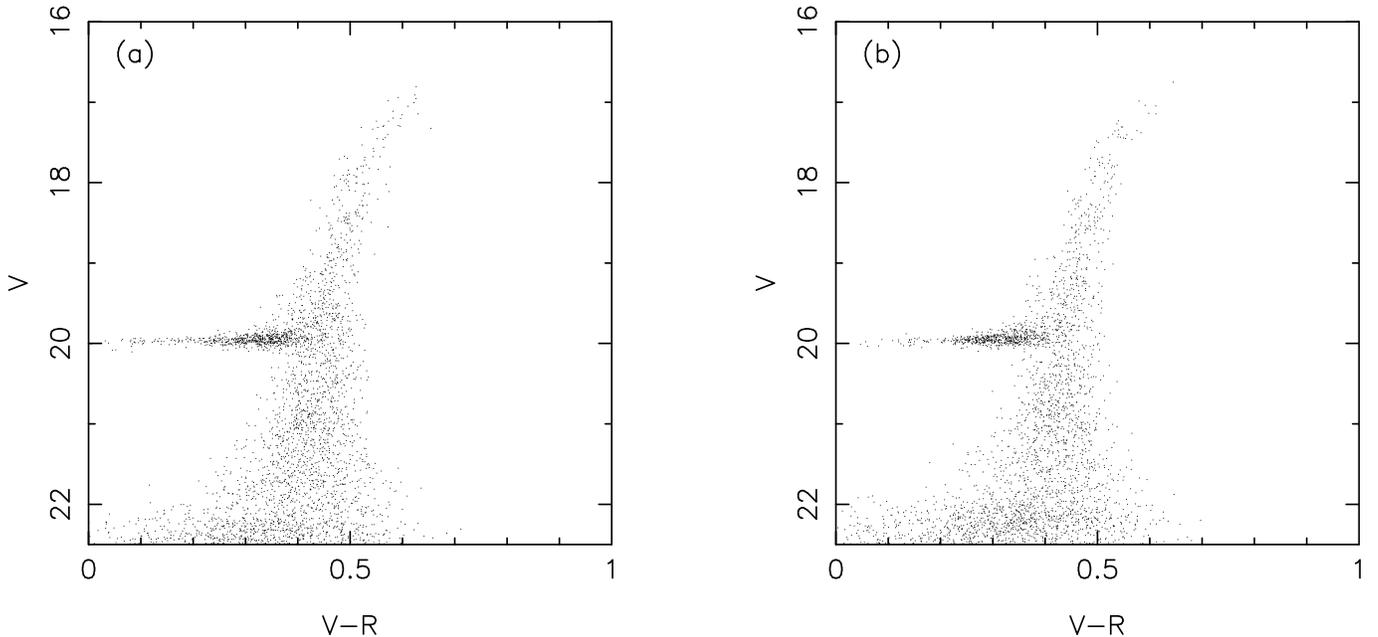}}}
\caption{The same as Fig.\,4, but for models F and G.}
\end{figure*}

Figs.\,4a and 4b represent the CMDs simulated  by 
models B and C, respectively, where the galactic age is assumed 
to be 12 Gyr, which is similar ages of Galactic globular clusters 
(e.g., Carretta et al. 2000);  
values of other parameters are written in table\,1. 
The photometric errors 
and detection completeness in the simulations are taken from 
Piatek et al. (2001). Both models can reproduce 
characteristics of the CMD of the dSph Draco, such as 
narrow RGB and heavily populated RHB.  
Despite the metallicity dispersion 
($-3 \le {\rm [Fe/H]} \le -1.4$), the theoretical 
CMDs results in narrow RGBs. 
This is due to the so-called age-metallicity degeneracy
as explained below.

Fig.\,5 shows CMDs simulated by assuming models B and C, 
where different colours represent 
different metallicity range as described in the figure. 
For presentation purpose, Figs.\,5a and 5b contain larger numbers 
of stars than those of Figs.\,4a and 4b, respectively.   
The adopted observational conditions are the same as those of Fig.\,4. 
Clearly, there is little correlation between the colour and 
metallicity of RGB stars because of the age-metallicity degeneracy. 
Since stars of higher metallicity are younger,  
the metallicity and age effects on the RGB colours are 
cancelled out.  This keeps the RGB narrow and tight. 
The age effect appears in the HB morphology too. 
Since only metal-poor and old ($>10$ Gyr) stars can evolve to 
BHB, younger core-He burning stars lie in the red part of the HB. 
The populous red HBs in Fig.\,5  clearly show  
this age effect on the HB morphology. 
A weak correlation between the stellar colours and metallicities 
is found in the bright ($m_V < 18$ or $M_V <-1$) RGB. 

In model C, the RHB is more populous and brighter than in model B. 
This is because the longer duration of the star formation period 
increases the number of younger and more massive core-He burning 
stars. However, the difference between models B and C is 
so small that it is difficult to determine the final epoch 
of star formation from the CMD that 
does not reach down to the turnoff level. 
Nevertheless, it can be said that, in the dSph Draco,  
chemical evolution was very slow and the duration of 
star formation was at least longer than 3.9$-6.5$ Gyr.

This evolutionary picture can explain both the narrow RGB and 
the populous red HB of the Draco dSph. Thus, the second parameter 
problem in the dSph Draco is solved for the first time by introducing  
relatively long duration of star formation ($>3.9-6.5$\,Gyr) 
which is fully consistent with the observed abundance patterns.

\section{Discussion}

The abundance patterns and the CMDs of the dSphs Draco, Sextans, 
and Ursa Minor share the common features. 
It can be said that they have similar SFHs and CEHs. 
The CMD of the Sextans dSph (see figure 2 in Suntzeff et al. 1993; 
figure 1 in Shetrone et al. 2001) is characterised by a narrow RGB and 
a red HB similar to the dSph Draco.  
This suggests that the SFH of the dSph Sextans is similar to that 
of the dSph Draco. In the CMD of the Ursa Minor dSph 
(see figure 1 in Shetorne et al. 2001), 
the blue HB is more populous and the RGB is narrower. 
This might suggest that the time scale of star formation 
was shorter in the dSph Ursa Minor.  
Nevertheless, it can be said that, in the Draco dSph,  
the SFR was very low and the duration of 
star formation was at least longer than 3.9$-6.5$ Gyr. 
For the other dSphs, it is safe to say that 
the characteristics of the SFHs are 
the low SFRs and the long durations ($>$ several Gyrs)  
of the star formation period. 

For the dSph Draco, model C may conflict with the CMD obtained 
by the Hubble Space Telescope (Grillmair et al. 1998).  
Through a comparison with the fiducial lines of metal-poor globular 
clusters M68 and M92,  Grillmair et al. (1998) suggested that the dSph Draco  
is older than  M68 and M92 by $1.6 \pm 2.5$ Gyr. 
Their result, however, seems inconclusive. Firstly, the 
sparseness of the photometric sample makes it difficult to measure 
the contribution of any intermediate-age population.  
Although a predominant RHB is the characteristic of the dSph Draco,  
the few RHB stars appeared in their CMDs. 
Secondly, they used different techniques for photometry 
and presented the combined results.  
For bright stars aperture photometry was adopted, 
while  the point spread function fitting method was 
used for faint stars. The border is unclear. 
This causes systematic discrepancies between 
the photometry of faint and bright stars. 
Third, a comparison with M68 and M92 is questionable. 
They used the F606W filter of the WFPC2 system. 
However, the WFPC2 Instrument Handbook 
recommends that the F555W filter is a better approximation 
to the Johnson V-band than the F606W. 
Their filter selection could have reduced the accuracy of 
the photometric calibration. 
Because of the combination of these uncertainties, 
it is dangerous to place much faith on their age estimation. 
More data are required to clarify these matters and  
images of a wide special coverage are particularly important.

\begin{figure*}
\resizebox{17.cm}{!}{\includegraphics{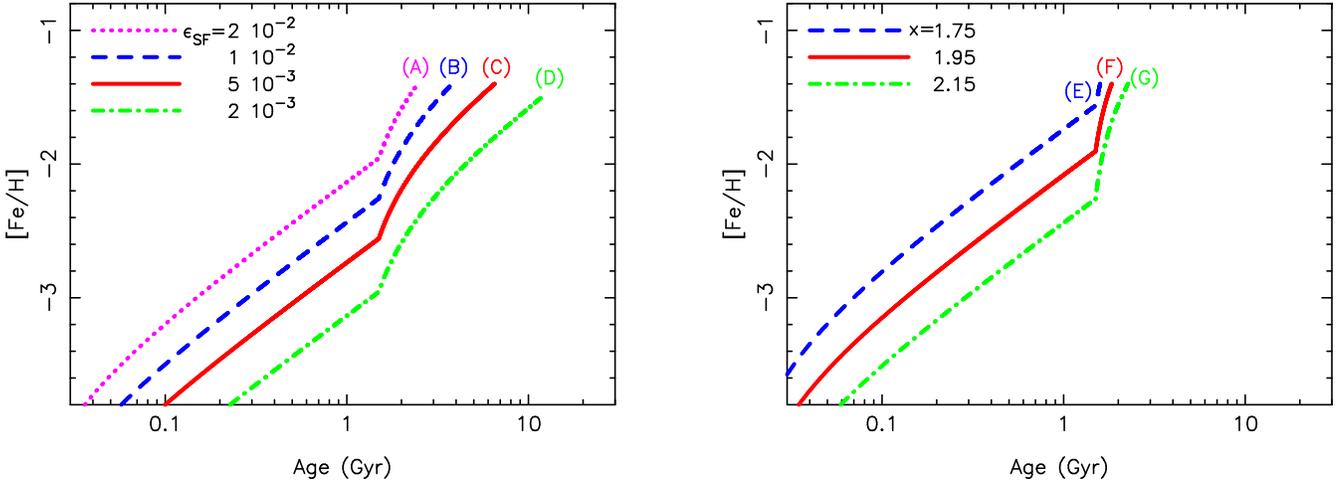}}
\caption{
Age-metallicity relations of models A$-$D (panel (a)) 
and E$-$G (panel (b)).  
The adopted values of 
$\epsilon_{\rm SF}$\,(Gyr$^{-1}$) 
or IMF power are shown on the panel.Colour version of 
this figure appears in the electric edition of the 
journal.}
\end{figure*}

In the above discussion, the decline in [Mg/Fe] observed in 
the dSphs is interpreted as the effect of the chemical enrichment by SNeIa. 
Since the signal-to-noise ratios of the data 
(Shetrone et al. 2001; S/N$\simeq 13-36$) for the dSphs 
are lower than other observations for metal-poor stars 
(e.g., McWilliams et al. 1995;S/N$\simeq30-40$),   
one might argue that the quality of the data 
is not high enough to study the CEHs. 
However, the goal of Shetrone et al. (2001) 
was the measurement of overall abundance differences, not 
absolute values as presented in their study,  
and it can be said that [$\alpha$/Fe] declines at [Fe/H]$\simeq -2$ on average 
(see figures 4 and 5 in Shetrone et al. 2001).
Thus, it is a pertinent interpretation that enrichment by SNeIa causes the 
decline of [$\alpha$/Fe].  
Spectroscopic observations of high signal-to-noise ratios are required 
to confirm this.

We have shown that low SFRs 
can explain the observed trends in the [Mg/Fe]$-$[Fe/H] diagram based on 
models assuming the Salpeter IMF. 
However, models of steeper IMFs and higher $\epsilon_{\rm SF}$ 
can also reproduce the trends. 
The steeper IMFs lead to slow chemical evolution, so that   
SNe\,Ia start to explode at low metallicity. 
This results in the decline of [Mg/Fe].

Fig.\,6 shows chemical evolution models E$-$G 
assuming different IMFs, where the IMF slopes are 
$x=1.75, 1.95,$ and $2.15$, respectively, 
and $\epsilon_{\rm SF}=0.2$ 
(a standard SFR for the solar neighbourhood; Arimoto et al. 1992).   
These models, except for model E,  are also consistent with the observations.
As with model A, the abundance ratios 
predicted by model E are too high and thus inconsistent 
with the observations. 
For the observed metallicity range ([Fe/H]$< -1.4$), 
the trends of [Mg/Fe] of models E$-$G are very similar  
to those of models A$-$D, although they become different 
at higher metallicity (see Fig.\,1).

Figs.\,7a and \,7b represent the CMDs simulated 
by models F and G, respectively. 
The photometric errors and detection completeness are 
taken from Piatek et al. (2001), i.e., 
the same as the simulations of models B and C shown in 
Fig.\,4. The CMDs in Fig.\,7 resemble those in 
figure\,4 in terms of the morphologies of RGB and HB. 
The slow chemical evolution and the long ($>1.6$ Gyr) 
durations of star formation of models F and G result in  
a narrow RGB and the populous RHB by the same reasons 
as those of models B and C.  This indicates that both abundance 
patterns and morphologies of the RGB and HB become similar when 
models have similar age-metallicity relations (AMRs).

Figs.\,8a and 8b show the AMRs of models A$-$D and 
E$-$G, respectively. The figures demonstrate that 
the AMRs of models E$-$G with the steeper IMFs 
result in very similar to those of models A$-$D with the lower SFRs  
at low-metallicity ([Fe/H]$<-1.4$).  Since both the abundance patterns 
and the morphologies of the RGB and HB are affected by CEHs, 
an additional information is necessary to solve the degeneracy of IMF and SFR. 
In the models \,E$-$G the initial star formation lasts 
only $1.6-2.2$\,Gyrs, while in the models \,B$-$C it continues 
as long as $3.9 - 6.5$\,Gyrs. Therefore, the models \,B$-$C should 
give systematically brighter turnoff magnitudes 
than the models\,E$-$G, and magnitude difference 
between the main sequence turnoff and the HB $V({\rm TO}-{\rm HB})$ 
in the models\,E$-$G should be smaller than in the models\,B$-$C. 
The models\,E$-$G give $V({\rm TO}-{\rm HB})$ is $\sim 3.1$\,mag, 
while the models B and C does $V({\rm TO}-{\rm HB}) \sim 3.1$\,mag 
and $\sim 2.8$\,mag, 
respectively. Deeper images with wider field of views 
will allow us to determine the final epochs of the initial star formation.

Figs.\,1 and 6 show that [Mg/Fe] converges at higher metallicity 
in the models with the different SFRs and that it does not
in those with the different IMFs. 
This difference allows us to derive the IMF and SFR independently. 
Abundance patterns in more massive dwarf galaxies and/or 
dwarf irregular galaxies will test whether the IMF or SFR produces 
in the abundance patterns and low metallicities in the dSphs. 
We note that even steeper ($x>2.75$) IMFs are required to 
explain the low metallicities of dwarf irregular galaxies if the IMF is the 
primary cause of the low metallicities. 
The CMD of the dwarf irregular galaxies in the Local Group reveal 
that they contain very old ($>10$\,Gyr) populations as well as young ones, 
i.e., the star formation and chemical enrichment have continued 
for at least 10\,Gyr.    Since they are still metal-poor, 
extremely steep ($x>2.75$) IMFs are required, which have never been reported. 
Therefore, we believe that low SFRs are the primary characteristic 
of the SFHs of the dSphs.

So far, the gas infall and outflow during star formation have been neglected, 
since no evidence of the gas infall and outflow has presented. 
Metallicity distribution could provide a clue  
to judge whether gas infall or outflow should be considered. 
We stress that spectroscopic observations is indispensable 
to derive a metallicity distribution. 
Fig.\,5 demonstrates little correlation between 
the colours and metallicities of the RGB stars. 
Because of the age-metallicity degeneracy and 
contamination by AGB stars, a stellar colour is not always 
a good indicator of the metallicity. 
Thus, a careful procedure should be adopted to convert 
stellar colours into metallicities (Ikuta 2001), although 
a relatively simple technique is often used (e.g., Harris \& Harris 1999).

Long lasting star formation at a very low rate 
explains the observed trend of [Mg/Fe] and the CMDs.  
Because of the low SFRs, plenty of gas 
($\sim 97$ percent of the galaxy mass) 
still remain even at the final epoch of star formation
(i.e., $t=\Delta T_{\rm SF}$). 
The remaining gas has to be removed to complete star formation and 
to evolve to a gas-poor system. Energy injection from supernovae   
could not be a sufficient mechanism to expel the gas 
from dwarf galaxies  (MacLow \& Ferrara 1999; Ikuta 2001). 
If this is the case, gas removal from dwarf galaxies should result from 
external mechanisms such as ram pressure stripping and/or 
tidal shocks. Studying the SFHs of 18 dwarf galaxies in the Local Group 
based on the CMDs derived by a uniform method of photometry, 
Ikuta (2001) found that the durations of star formation period 
correlate with the distances from the Milky Way or M31.  
The correlation may suggest that the environmental effects play a key 
role in the evolution of the dwarf galaxies.
Numerical simulations (Moore et al. 1998; Mayer et al. 2001) clearly 
showed that late-type dwarf galaxies entering the dark matter 
halo of a massive galaxy are transformed into 
early-type owing to repeated tidal stripping 
and dynamical instabilities. 
Thus, we conclude that the star formation 
in the dSphs Draco, Sextans, and Ursa Minor 
terminated due to gas stripping by the Milky Way.

The hierarchical clustering galaxy formation model 
suggested that star formation in dwarf galaxies can only occur 
before the cosmological re-ionisation epoch and there is no 
major star formation activities later because gas cannot cool and 
condense to form stars (e.g., Cen 2001).  
This appears to be inconsistent with 
stellar populations observed in the Local Group dwarf galaxies. 
CMDs obtained by the HST (e.g., Ikuta 2001) and 
our study presented in this paper clearly show  
extended and recent star formation activities in the Local Group 
dwarf galaxies.The issue is still being debated.   
However, a more recent simulation (Kitayama et al. 2001) 
suggested that star formation can occur in 
small objects if they have baryonic mass larger than the threshold mass 
of $10^9 M_\odot$ and $10^6 M_\odot$ at redshifts 
of $< 3$ and $\sim 5$, respectively. 
The masses of today's dSphs in the Local Group 
are $10^7 - 10^8 M_\odot$ (e.g., Mateo 1998) and  
our results imply that progenitors of today's dSphs lost 
$\sim 97$\,\% of their mass after the long lasting star formation.  
Therefore, the masses of the progenitors are estimated at  
$10^9 - 10^{10} M_\odot$. 
Since these exceed the threshold predicted by Kitayama et al. (2001),  
the SFHs derived here do not contradict 
recent simulations of the formation of dwarf galaxies.

\section{Conclusion}

Based on comparisons between the theoretical 
chemical evolution models and the observed abundance patterns,  
we conclude that the initial star formation continued 
for a long duration ($> 3.9-6.5$ Gyr) in the dSphs Draco, Sextans, 
and Ursa Minor. Our simulation of the CMDs shows that 
the long duration of star formation can solve  
the second parameter problem of the Draco dSph. 
Because of the age-metallicity degeneracy, 
the RGBs are kept narrow and tight despite their large metallicity 
dispersions.

We have discussed the two cases which are consistent with the observed 
abundance patterns and the CMDs.
The first case is a combination of  low SFRs 
($\epsilon_{\rm SF}=1 \cdot 10^{-2} - 5 \cdot 10^{-3}$ Gyr$^{-1}$) 
and the Salpeter IMF ($x=1.35$), while 
the second is a combination of the solar neighbourhood SFR ($\epsilon_{\rm SF}=
0.2$ Gyr$^{-1}$) and the steeper IMFs ($x=1.75-2.15$). 
The two cases are discriminated neither by the abundance 
patterns nor by the CMDs of bright stars. 
However, the initial star formation period is as long as $3.9-6.5$\,Gyrs
if the SFR is low, while it is much shorter ($1.6-2.2$\,Gyrs) if the 
SFR is high. Thus, turnoff magnitude should be different between 
the two cases. Deeper images with wider field of views 
allow us to determine the final epochs of the initial star formation. 

\begin{acknowledgements}
We wish to express my gratitude to the anonymous referee 
for very helpful suggestions and comments. 
We are also grateful to T. Kodama, and H. Susa, 
for fruitful discussion and comments and  
to B. Jones who carefully read this manuscript and 
gave us helpful comments and suggestions. 
C.I. wishes to thank  
the Japan Society for Promotion of Science for financial support. 
\end{acknowledgements}

\clearpage

\end{document}